\documentclass{ws-rv9x6}
\usepackage{subfigure}   % required only when side-by-side / subfigures are used
\usepackage{ws-rv-thm}   % comment this line when `amsthm / theorem / ntheorem` package is used
\usepackage{ws-rv-van}   % numbered citation & references (default)
\makeindex
\begin{document}

\chapter[2]{Computing with Clocks}
\author[]{Jonathan Edwards, Alex Yakovlev, Simon O'Keefe}
\address{University of York, Newcastle University}

\begin{abstract}

Clocks are a central part of many computing paradigms, and are mainly used to synchronise the delicate operation of switching, necessary to drive modern computational processes. Unfortunately, this synchronisation process is reaching a natural ``apocalypse''. No longer can clock scaling be used as a blunt tool to accelerate computation, we are up against the natural limits of switching and synchronisation across large processors. Therefore, we need to rethink how time is utilised in computation, using it more naturally in the role of representing \textit{data}. This can be achieved by using a time interval delineated by discrete start and end events, and by re-casting computational operations into the time domain. With this, computer systems can be developed that are naturally scaleable in time and space, and can use ambient time references built to the best effort of the available technology.

Our ambition is to better manage the energy/computation time trade-off, and to explicitly embed the resolution of the data in the time domain. We aim to recast calculations into the ``for free'' format that time offers, and in addition, perform these calculations at the highest clock or oscillator resolution possible.
\end{abstract}

\section{Introduction}\label{intro}

We are at the computational apocalypse. The rate of increase in the ability to compute is asymptotic \cite{thompson2020computational}, and we are collectively hiding from the harsh reality that we need to rethink how to deliver computation for global productivity and the planet. This is because much of the increase in productivity in the medium term is pinned on the replacement of knowledge work, and hence workers, with AI systems. We hope to work more efficiently with knowledge and information, through advancing the range of activities we can leave a computer to do, unsupervised by a human.

An example of this is the British National Health Service (NHS). It is currently a two stage system of knowledge distribution that means the first line (primary care) acts like a filter for more detailed knowledge at the later stage (secondary care). Currently, primary care is deluged with health complaints, which means they only provide a sparing service. In a utopian future this filter could be scaled and automated by computers. This will not only increase the throughput of patients but, since we can then access secondary care knowledge as easily as primary care, also increase the quality of service. This could be a really big deal for society. 

But what of these AI systems? Our current instantiation involves what we presently consider ``Deep'' learning systems \cite{lecun2015deep}, where the training algorithms have to be scaled in such a way that we reach practical physical limitations. We need too many processors, too much memory, too much energy, and this all occupies too much physical space. Furthermore, Moore's law \cite{Moore:2000:CMC:333067.333074} is failing this system, as we have an exponentially increased need, against a backdrop of only modest improvements in Floating Point Operations Per Second (FLOPS) and FLOPS per WATT \cite{moorefail}. Currently (2023), only a few organisations on the planet have enough resources to plan to build these systems. This naturally forms a new oligopoly, perpetuating the control these organisations have over society. So, the \textit{actual} impact of Moore's Law failing is a failure in the democratisation of computation, the ``haves'' increasing the hold their technology has over the main digital interfaces in society, eg. the cloud and the internet, the desktop and laptop computer, and the mobile phone. 

So what, realistically, can we do about our methods? Clearly, we cannot continue in the same vein. Are we so heavily invested in the miniaturisation of synchronous and somewhat parallel digital binary workloads, that we cannot think of alternatives even though we see the dead end in sight? To use a transport analogy, this reluctance to fundamentally change course means we will have supremely fast horses, with little or no other modes of transport for the increasing variety of journeys (workloads) which require delivery.

There \textit{are} other technologies available, but even then we are still yet to liberate computation from it’s mainly digital, wholly logic-gate basis. Quantum computing \cite{aaronson2013quantum} (a vehicle that solves niche problems, something like a helicopter!) is a great hope, but there is some doubt that this can be utilised in the correct way to solve more general computation problems, rather than just those specific to complex combinatorial optimisation \cite{feynman2018simulating}.

There is a completely overlooked area of computation that has the potential to liberate us from the hold of Digital Binary Synchronous (DBS) systems. This technology we will name as Temporal/Unary/Asynchronous or \textbf{Temporal Computing}, and in this article we will explain this in more detail, discussing possible implementation methods and summarising our agenda for making it work, and hence the world fundamentally more democratic from a computational perspective.

Our long-term research and development agenda is as follows:
\begin{enumerate}
    \item Clocks to become the main computation fabric.
    \item Unary to outperform binary arithmetic.
    \item Co-ordination to become asynchronous (and heavily parallel).   
\end{enumerate}

In the following sections we will discuss each of these features in isolation before we detail how they all might fit into a cohesive whole. Our aims are more of a practical nature, rather than an exploration of the theoretical properties.

\section{Clocks and Accumulators}

The key new idea in temporal computing is that input data is represented as a \textit{time-delay}. Typically, this may be the time taken for some noticeable change to happen, in either an analog or digital signal, and is delineated by start and end markers. Our reliance on time delays is motivated because of the following:

\begin{enumerate}
    \item Time is a free resource \cite{timeguy} and ordered sequentially.
    \item Clocks can run faster than current computer frequencies, and can be set to run very fast \cite{nicholson2015systematic}.
    \item Very general computing in unary can be effectively organised into a temporal system.
    \item There are many physical media that can support temporal computation methods \cite{csaba2020coupled}.
    \item Computation speed is naturally linked to the size of the data.
\end{enumerate}

\begin{figure}
\centerline{\includegraphics[width=9cm]{"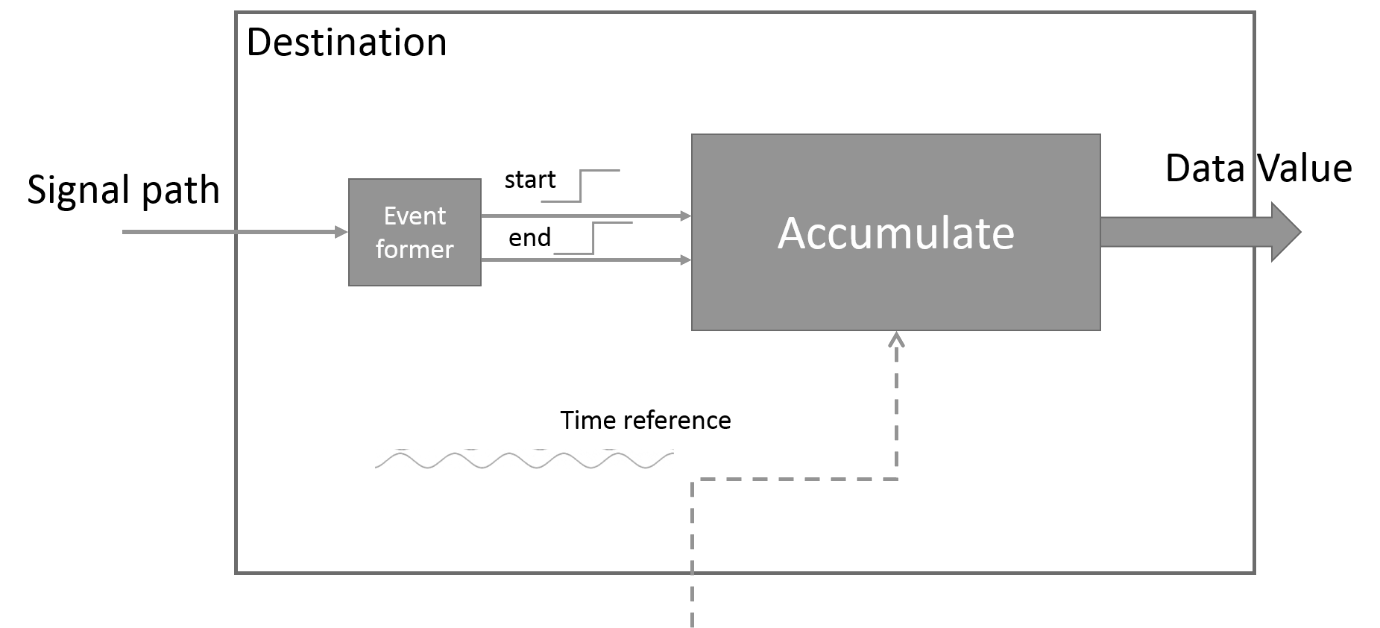"}}
\caption{Time as a complementary channel, for an example temporal computation that utilises accumulation.} \label{im3}
\end{figure}

To describe the system more formally, we need to refer to time as a separate and complementary channel, in addition to the carrier channel, i.e. a channel that isn't \textit{directly} carrying the signal but is never-the-less integral in the computation. Then, by definition, temporal computation is any system regardless of the composition of the main channel, that uses this complementary time channel to perform computation. This is shown in Figure \ref{im3}, we notionally have a accumulation unit (an adder)  which takes both a set of electrical signals \textit{and} a clock, to perform its computation. 

The most basic arrangement for the time channel is that it counts in a linear fashion and hence gives an underlying representational scheme based on unary and unary coding schemes, which we discuss in Section \ref{unary}. 

\begin{figure}
\centerline{\includegraphics[width=7.8cm]{"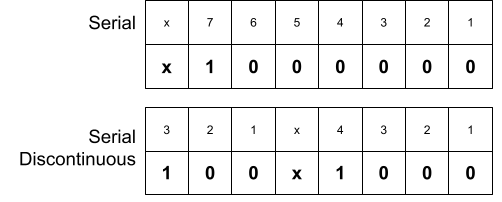"}}
\caption{Sequential temporal systems. There are two possible systems, Serial and Serial Discontinuous.
We have made the time axis discrete and starting from the right going leftwards, how you might write a number conventionally. The bases I have used throughout are unary and an \texttt{x} mark indicates an absence of signal. To read the diagram move from the right to the left. } \label{s1}
\end{figure}

\begin{figure}
\centerline{\includegraphics[width=8.8cm]{"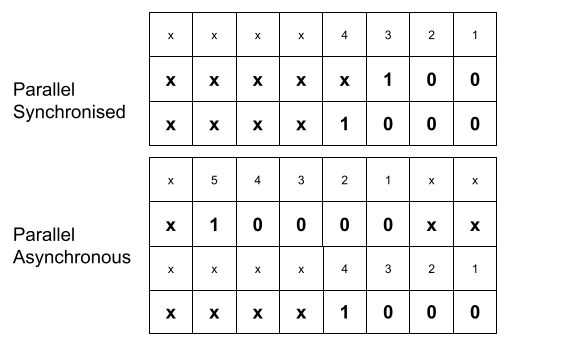"}}
\caption{Parallel temporal systems. There are two possible systems, Parallel Synchronous and Parallel Asynchronous.} \label{p1}
\end{figure}

Further to this, there are a variety of ways one might organise the time delays, and this forms a taxonomy of possible temporal computation approaches (see Figures \ref{s1} and \ref{p1}). 
The taxonomy can be basically split into whether the time delays are supplied in a serial or parallel way. For serial systems, we will see that for some operations (most obviously \textit{add}) it is advantageous for the time delays to form a continuous stream, but there are also sequential systems that have discontinuities between time delays (Figure \ref{s1}). Parallel temporal systems also come in two forms - the first and again the most advantageous, is where the time delays are delivered from a synchronous start point and likewise there are systems where the time delays arrive at any time (Figure \ref{p1}).

The above definition focuses  on time delays and the complementary timing channel, without saying much practically about the type of media that can be used to build a temporal computer. So far very few temporal systems have been built - the most notable are those of Madhaven \cite{madhavan2015race} and Wu \cite{wu2020ugemm}. These use simple digital logic-gates to perform \texttt{min} and \texttt{max} operation which \textit{are} measured using a clocking channel. 
Although we fixate on clocks, in practice there is only a loose requirement that a clock should be involved, any media that changes over time can be deployed. Below is a series of descriptions of temporal systems based on Digital, Analog and Photonic technologies. 

\subsection{Digital} 

Digital circuits involve a traditional clock where the oscillator is physical, like a quartz crystal - something with a consistent natural vibratory property, or the oscillator is a stable dynamic system such as a ring oscillator. One is purely digital – it is a counter that counts time reference pulses trapped between the start and end signals. 
There are many ways of implementing such a counter using asynchronous circuit design techniques \cite{yakovlev2000hardware}. If the counter is built as a composition of toggle latches, where each toggle latch divides the frequency of its input signal by 2, the accumulated value will be in a binary code, whose size is logarithmic to the maximum data value. 

If the time reference signal is a conventional clock (from Phase Locked Loop (PLL) or Voltage Controlled Oscillator (VCO)) \cite{ring}, as a sensing element we can use an MOS transistor, which will act as a current switch modulated by the clock signal connected to its gate. 

\subsection{Analog} 

\begin{figure}
\centerline{\includegraphics[width=9.8cm]{"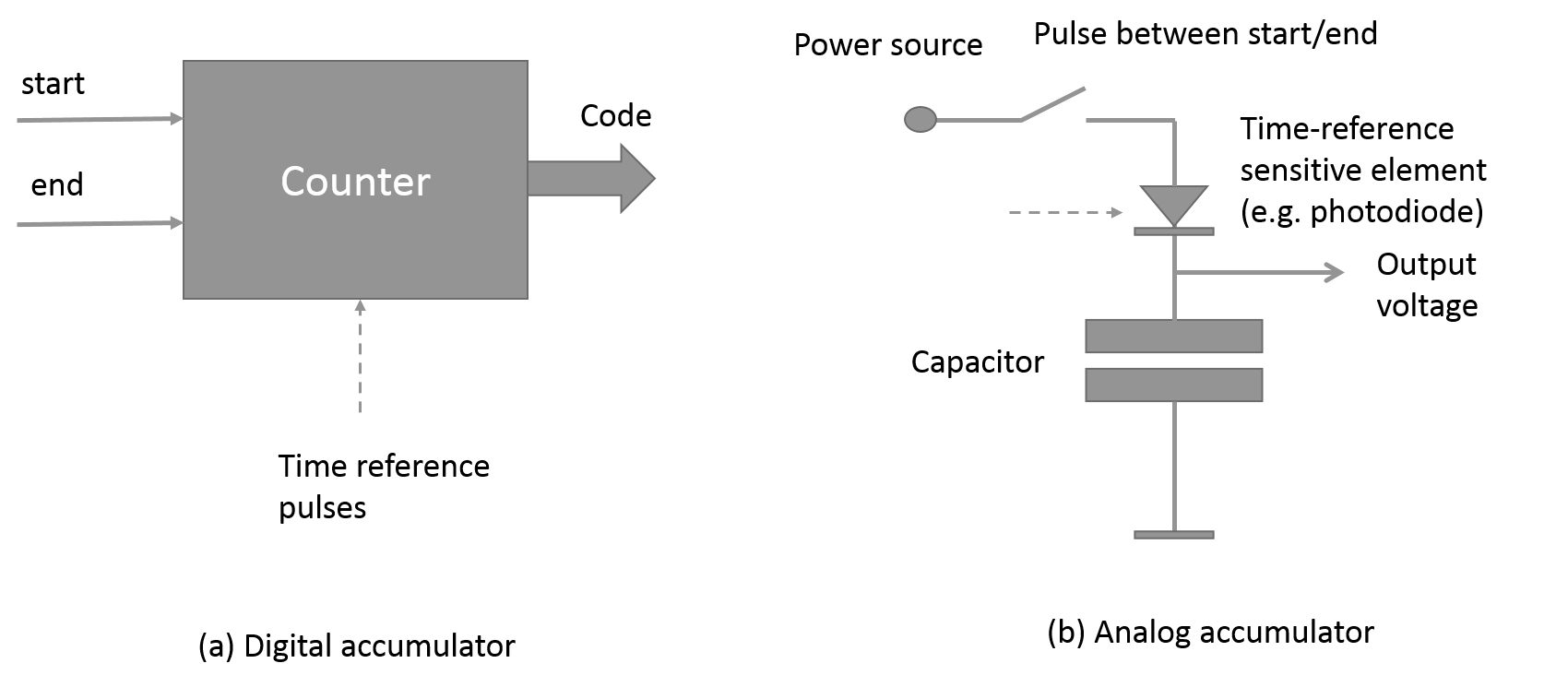"}}
\caption{Simple mechanisms to perform temporal computation in digital, analog and photonic.} \label{accum}
\end{figure}

Another implementation of a temporal system could be analog, and rely heavily on \textit{accumulation}, for example by means of a capacitor. Here, and utilising the start and end events shown in Figure \ref{im3}, we also need a voltage source, a switch which is controlled by the received interval (the switch is \texttt{ON} during the interval), and an element that is activated by the time reference signal and conducts the charging current at the rate related (e.g. proportional) to the frequency of the time reference. The result of this accumulation is produced as a data value for the needs of subsequent computation inside the destination block.  Figure \ref{accum} illustrates two possible ways of realising the Accumulate unit. 

Taking the idea of an analog accumulator further, to much higher resolution levels, we can envisage an accumulator being a device built of materials sensitive to other types of radiation, such as X-rays or gamma-rays.

\subsection{Photonics} 

This could again involve accumulations but through a Charge Coupled Device (CCD) that collects photons. This can be built using a photo-diode if the time reference is light of a particular frequency. The total charge accumulated by the device, and hence the output voltage (which is the data value), will be proportional to the length of the received time interval. 

Now that we have formalised the medium we might use to compute with time-delays, we need to discuss the basic unary strategies for performing arithmetic.

\section{Unary and Unary Codes} \label{unary}

\begin{figure}
\centerline{\includegraphics[width=7.8cm]{"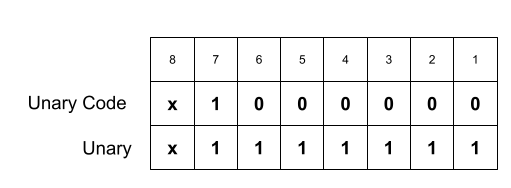"}}
\caption{Unary and Unary Codes.} \label{unarypic}
\end{figure}

Unary number systems (Mackay \cite{MacKay:2002:ITI:971143} Chapter 7 page 133) were amongst the earliest numerical representation of quantity, with the abacus existing as the earliest calculating device \cite{POPPELBAUM198747}. Unary codes are a simplification, with a leading \texttt{1} acting as the delimiter. In abstraction, unary codes can be seen as an \textit{interval code} where the number is represented as a pulse followed by a non-signal. Interval codes have been exploited in the communication community in the form of Pulse Interval modulation (PIM) \cite{ghassemlooy2000digital}, where the pulses delimit both the beginning and end of the coding. Figure \ref{unarypic} shows an example coding for the value 7.

The basic model holds that length, either explicitly or logarithmically scaled, acts as the numerical representation. This idea, for temporal representations, was first mentioned in the work of MacKay and McCullough \cite{MacKay1952} who focused on the channel rates in idealised biological neurons. This work was extended as \textbf{Action Potentials}, with significant work by Hopfield  \cite{hopfield1995pattern} and Maass \cite{maass1997networks}. From this the ``neuromorphic''community  has emerged, fuelling more recent production of commercial hardware by companies including IBM \cite{ibm} and Brainchip \cite{brainchip}.

\subsection{Adding Unary Codes}

\begin{figure}
\centerline{\includegraphics[width=7.8cm]{"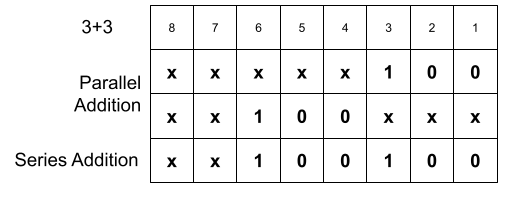"}}
\caption{Addition via concatenation.} \label{add}
\end{figure}

Addition with unary, as with the abacus, is trivially performed using simple concatenation as shown in Figure \ref{add}. The benefit of this approach is that the representation does most of the problem solving, we are computing \textit{naturally} because we have represented the data in the right format.

\subsection{Multiplying Unary Codes}

\begin{figure}
\centerline{\includegraphics[width=7.8cm]{"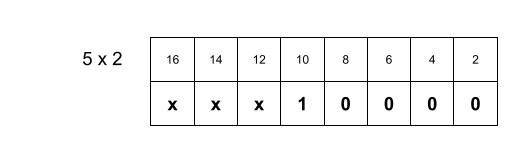"}}
\caption{Multiplication by changing the clock frequency.} \label{mul}
\end{figure}

Integer multiplication of unary codes is performed by dilating the first unary code by size of second operand. This is particularly useful in the clocked domain since each pulse can be scaled by changing the clock frequency to effect the dilation. Again, this is straightforward because we are representing numbers as lengths which can be easily scaled, this is shown in Figure \ref{mul}. 

\subsection{Hybrid Positional Schemes}

The major criticism of unary is the linear scaling of the representation size with the value stored. Position based coding has at it’s heart a compression property that provides invaluable log scaling for storage capacity which we are accustomed to in the digital world, together with the more problematic \textit{non-self-delimiting} bit-widths. There are other schemes which utilise positional significance, most notably Binary Coded Decimal (BCD) \cite{arch}. It is straightforward then, to imagine many combinations of position and unary systems, to exploit the computational simplicity of unary codes and mitigate somewhat the comparative memory inefficiencies when compared to binary. Of course this memory efficiency is only relevant if we are thinking about memory in a traditional setting.

\subsection{Multiplexed Unary}

\begin{figure}
\centerline{\includegraphics[width=7.8cm]{"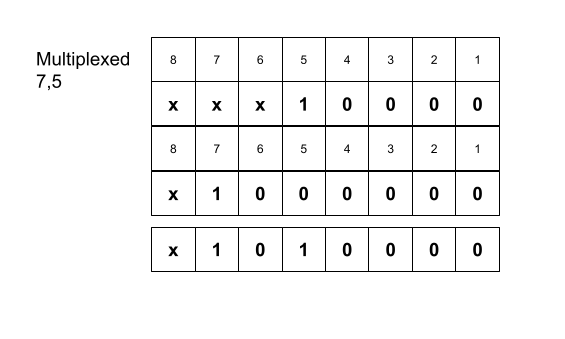"}}
\caption{Two values are transmitted in the time interval between T=0 and T=7. Clearly the intervals up to T=5 are used in the transmission of the signal at 7.} \label{plex}
\end{figure}

Unary codes have the additional property that a set of non-duplicate values can be efficiently multiplexed. Again, to use the abacus analogy, if we have beads of two colours (black (1) and white (0)) then provided that there is no duplication, multiplexing can occur which \textit{reuses white beads}. Again, the logic of this is a simple `OR'ing of the value from a start reference bit. An example of this multiplexing is shown in Figure \ref{plex}, which shows one channel with the values 5 and 7. The two values are genuinely overlaid so the value 5 actually forms part of the data for the value 7. This is unique to this coding strategy and is unlike any other system of multiplexing.

\subsection{Multi-valent Computation}

\begin{figure}
\centerline{\includegraphics[width=7.8cm]{"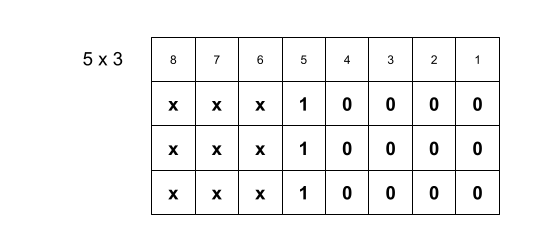"}}
\caption{A multi-valent $5\times3$ in a temporal notation.} \label{val}
\end{figure}

A natural extension is to enable the unary coded \texttt{1} value to be multi-valent. Although at first this may appear contrived, the motivation for this is the extension of the multiplexing method to incorporate duplicates, and the ability to better encode analog interval channels. This channel also has an additional property that multiplication can be performed, at least within the bounds of the notation, by placing a value $a$ at ``temporal'' position $b$, for tuple $a \times b$. Figure \ref{val} shows 5×3 with a multi-valent 3.

\begin{figure}
\centerline{\includegraphics[width=7.8cm]{"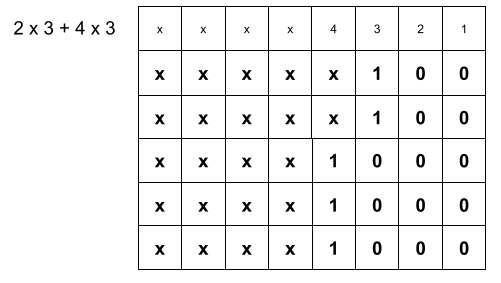"}}
\caption{A more complex temporal arrangement, here representing the dot product $2 \times 3 + 3 \times 4$. The answer can be evaluated using the MADD algorithm which efficiently counts the 1's and 0's. } \label{madd}
\end{figure}

Clearly, the act of multiplication can be transformed into manipulating the values held in the buckets. This can be easily extended to a dot-product operation and the representation of tuple pairs is somewhat compressive since one element of the 2-tuple is specified by a positional index. To illustrate, Figure \ref{madd} details the example $2 \times 3 + 3 \times 4$. A dot product on this multiplexed array can be evaluated efficiently and in a systolic way using the \textbf{Multiplicative ADD} (MADD) algorithm \cite{jonny-asynch}. 

\section{Asynchronous Control}

The key point behind any computation is not so much how we perform individual operations on data (these can be performed by means of the technology at hand), but how we coordinate computational processes between operators or blocks.

\begin{figure}
\centerline{\includegraphics[width=5.8cm]{"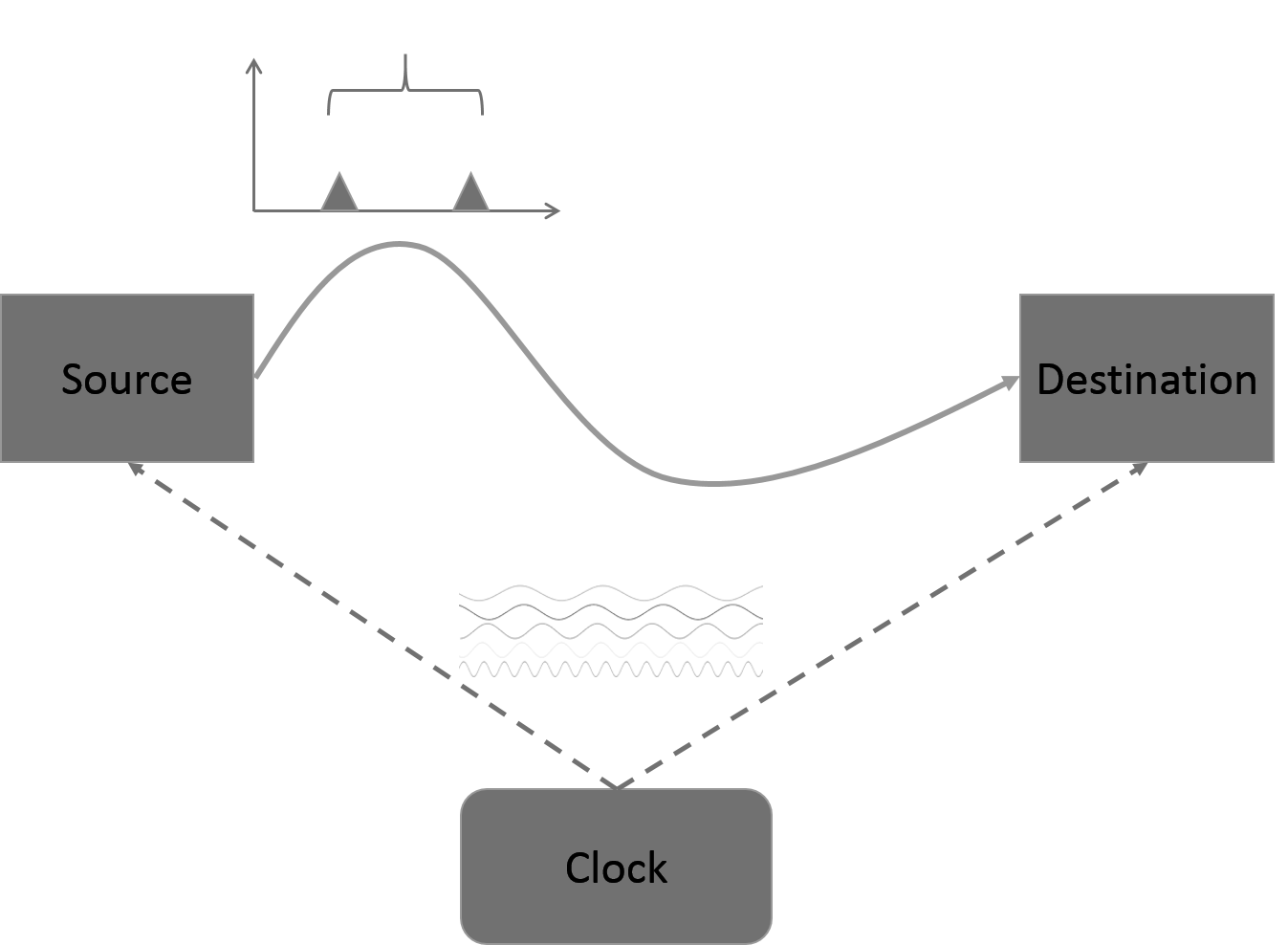"}}
\caption{How blocks communicate asynchronously.} \label{asy}
\end{figure}

Our method allows one to separate the way data is passed in the system between computing elements from the metrology of the data. There is no need to synchronise signalling control events with the clock. All signalling associated with data transfer is asynchronous. The information is passed between blocks independently of the propagation delays of the links and paths. The technology in which the communication fabric is designed will of course determine the limits of possible data precision, but the actual data resolution limits are determined by the time reference source and the ability of the communication fabric to sense and react to the time reference signals in a proportional (accumulating) way. Because the clock is not used for control purposes but only for measuring data, we can potentially use the clock at super-high frequency as long as our accumulator circuits maintain this sensitivity and reactiveness in a naturally averaging way.

In other words, it is about how to communicate information from one block to another. The basic idea of communication is illustrated in Figure \ref{asy}. The communication paths between source and destination will carry events and the intervals of time between the pairs of events will define data values. In what ‘currency’ those values are measured will be determined by the time reference (clock!) signal that will be agreed by both sides. What is important is that the length of the interval of time as the interval travels from the source to destination will have to remain unchanged. We call it the requirement of stability of the delay within the data interval. In some sense this is the requirement of the stability of the speed of light in the medium! This is likely to be satisfied under fairly reasonable assumptions. But, what is also crucial, is that the correctness of the data transfer will no longer depend on the latency of the transmission, and hence the route via which the timed-data interval was sent.

Another important requirement is that both ends of the communication path receive the same time reference during the period of communication. This requirement will effectively guarantee the integrity of the denomination of the ‘currency’ by which we count data on both sides. Figure \ref{asy} shows a range of possible time references, and clearly both sides need to be in agreement on which of them they choose to use. Actually, if there is a definite relationship between time references, such as definite frequency ratio (cf. currency exchange rate) the source and destination may work from those, even different, references. Such an option may be needed, for example, if the sides of the channel are implemented in different technologies.

\section{Summary}
In this work we take a step back from the computational models predominant binary fixation to address methods that only require manipulation of time-delays rather than electrical transistors. This may at first appear to be a retrograde step however unary has been shown to be a viable computational medium when applied in the time domain with several benefits, particularly when multiplexed.
We have spelled out a limited number of operations, but this sets a precedent for our major aim, which is to invent an entirely new computational model extending clock usage to the data representation, and using asynchronous methods instead for coordination.
The agenda of this work is quite clear--we aim to use the clock past its current limitations, in the belief that this may avoid the eventual stagnation of clock usage for computation. 

Our future plans are to extend the number of operations across a range of working media together with more formal definitions of the computational power of any of the formulations we produce. As a first ``port of call'' we aim for a ten fold increase in clock speed in performing a bit-width unlimited add operation in a purely temporal medium.

\bibliographystyle{ws-rv-van}
\bibliography{temporal}

\end{document}